\documentclass[a4paper,11pt]{JHEP3}
\usepackage{fancyhdr}
\usepackage[dvips]{graphicx}
\usepackage{ifthen}
\usepackage{amsmath}
\usepackage{epsfig}
\usepackage[errorshow]{tracefnt}
\newcommand\nn{\nonumber}

\newcommand\beal{\begin{align}}
\newcommand\eeal{\end{align}}
\newcommand\benu{\begin{enumerate}}
\newcommand\eenu{\end{enumerate}}
\newcommand\bit{\begin{itemize}}
\newcommand\eit{\end{itemize}}

\newcommand{\be}{\begin{equation}}
\newcommand{\la}{\label}
\newcommand{\ee}{\end{equation}}
\newcommand{\bd}{\begin{displaymath}}
\newcommand{\ed}{\end{displaymath}}

\newcommand\g{\gamma}
\newcommand\de{\delta}

\newcommand\e{\epsilon}

\newcommand\G{\Gamma}
\newcommand\Om{\Omega}

\newcommand\wh{\widetilde{h}}
\newcommand\wb{\widetilde{b}}
\newcommand\wg{\widetilde{g}}
\newcommand\wj{\widetilde{J}}

\title{${\cal N}=1,2$ supersymmetric vacua of IIA supergravity \\ 
and $SU(2)$ structures}

\author{Jo Bovy \\
Institute for Theoretical Physics, K.U.~Leuven, \\
Celestijnenlaan 200D, B-3001 Leuven, Belgium\\
E-mail: \email{Jo.Bovy@student.kuleuven.ac.be}
}
\author{Dieter L\"{u}st \\
Max-Planck-Institut f\"{u}r Physik --Theorie\\
F\"{o}hringer Ring 6, 80805 M\"{u}nchen, Germany\\ \&\\
Arnold-Sommerfeld-Center for Theoretical Physics\\
Department f\"{u}r Physik, Ludwig-Maximillians-Universit\"{a}t\\
Theresienstr. 37, 80333 M\"{u}nchen, Germany 
\\E-mail: \email{luest@mppmu.mpg.de, luest@theorie.physik.uni-muenchen.de}
}
\author{Dimitrios Tsimpis \\
Max-Planck-Institut f\"{u}r Physik --Theorie\\
F\"{o}hringer Ring 6,  80805 M\"{u}nchen, Germany\\
E-mail: \email{tsimpis@mppmu.mpg.de}
}

\abstract{ We consider backgrounds of (massive) IIA supergravity 
of the form of a warped product $M_{1,3}\times_{\omega} X_6$, where $X_6$ 
is a six-dimensional compact manifold and 
$M_{1,3}$ is $AdS_4$ or 
a four-dimensional Minkowski space. We analyse
 conditions for $\mathcal{N}=1$ and $\mathcal{N}=2$ supersymmetry 
on manifolds of $SU(2)$ structure. We prove the absence of solutions in  
certain cases.}

\preprint{MPP-2005-63, LMU-ASC 46/05}

\begin{document}

\section{Introduction and summary}

String theory compactifications in the presence of fluxes 
possess a number of phenomenologically attractive features,  
a fact which has led to their intensive study in recent years. String 
theory is approximated in the low-energy effective-theory 
limit by supergravity and in many situations 
of physical interest it has proven extremely fruitful to 
study the properties of supersymmetric supergravity solutions. 
The presence of some amount of unbroken supersymmetry above a certain  
low (compared to the Planck mass) energy scale is desirable,  
if one wishes to avoid issues of stability. 

The subject of supersymmetric supergravity compactifications 
with fluxes is a particularly old one. 
Nevertheless, it has only recently been realized 
(starting with \cite{gaun}; see \cite{gauntb} 
for a recent review and references)  
that the machinery of $G$-structures 
can be a powerful tool in classifying and 
constructing supergravity solutions. 
From this point of view, a 
$G$-structure is the natural generalization 
of the concept of special holonomy to the case where 
nontrivial fluxes, i.e. nonzero vevs of the antisymmetric tensor fields, 
are present.

In \cite{lt} we presented a 
classification of $\mathcal{N}=1$ supersymmetric solutions of 
IIA supergravity of the form 
of a warped product $AdS_4\times_{\omega} X_6$, where $X_6$ 
is a six-dimensional compact manifold of $SU(3)$ structure, generalizing 
the work of \cite{bca, bc}. The manifold $X_6$ was constrained to be 
`half-flat' of a certain type. For further related work on 
IIA compactifications from the point of view 
of the four-dimensional 
effective field theory see \cite{g1, g2, g3, g31, g4, g5, granb}. 
Type IIA compactifications have also been considered 
in the context  
 of $G$-structures in \cite{Cardoso:2002hd, gg1, gg2, gg3, gg4}. 
The recent paper \cite{grana} analyzes 
type II $\mathcal{N}=1$ supersymmetry using the concept of 
generalized $G$-structures --we will come back to this  
in the next  
paragraph.  
Supersymmetric $AdS_4$ solutions 
are of additional interest as they are expected to be dual to certain 
three-dimensional superconformal field theories  \cite{js, nunez}.

As will be explained in the following, 
for the backgrounds considered here
$\mathcal{N}=1$ supersymmetry 
implies that the Majorana-Weyl supersymmetry parameter $\e$
is of the form 
\beal
\e&=\theta_+\otimes(\alpha\eta_{1+}+\delta\eta_{2-})+{\rm c.c.}~,
\label{1}
\end{align}
where $\alpha$, $\delta$, are functions on $X_6$, $\eta_{1+}$ 
($\eta_{2-}$) is a globally-defined, chiral (antichiral)
 unimodular spinor (and therefore 
{\it nowhere-vanishing}) on $X_6$ and $\theta$ is a Killing spinor 
of $M_{1,3}$\footnote{$\eta_{1,2+}$ are related 
to $\eta_{1,2-}$ by complex conjugation. Equation (\ref{1}) represents 
one linear  combination of $\eta_{1,2}$
and corresponds to  $\mathcal{N}=1$ in $d=4$. }. 
The existence of $\eta_{1}$ 
implies that the structure group of $X_6$ is reduced to $SU(3)$. 
If in addition 
$\eta_{1,2}$ are nowhere-parallel, the structure group 
is further reduced to $SU(2)$. 
Relaxing the condition that $\eta_{1,2}$ should be linearly-independent
everywhere on $X_6$ would lead to a situation which can be thought of
as a so-called `generalized $SU(3)$ structure' on $X_6$
\cite{jesc, gg3,  kapu, jescb, grana, granb}: 
at generic points
in $X_6$ the two $SU(3)$ structures associated 
with each of the two internal spinors
have a common subgroup, which defines
an $SU(2)$ structure on $X_6$. However, at the points where the two spinors
become parallel the structure group collapses to $SU(3)$. 

Supersymmetric vacua on manifolds of $SU(2)$ structure restrict the choice
for the fluxes similar to their  $SU(3)$-structure 
counterparts. In addition, the requirement of an $SU(2)$ structure 
imposes a strong constraint on the internal 
manifold\footnote{A necessary and sufficient 
condition for the structure group of 
a manifold $X_6$ of $SU(3)$ structure to 
further reduce to $SU(2)$, is the existence of a
globally-defined nowhere-vanishing 
vector field on $X_6$. This is equivalent 
to the requirement that the Euler
characteristic vanish, $\chi(X_6)=0$.} and one may hope 
that a classification can proceed much more explicitly in this case. 
However the situation is much more difficult to analyze 
in practice, and this subject has received 
much less attention in the literature, 
because of the multitude of flux-components 
which arise in decomposing the supergravity fields 
in terms of irreducible $SU(2)$ representations.

In the present paper we examine 
$\mathcal{N}=1$ supersymmetric type IIA vacua 
in the case where $X_6$ is a 
compact manifold of $SU(2)$ structure. 
In \cite{lt} we considered the case $\eta_{1}=\eta_{2}$. Here 
we will consider the other `extreme' case where $\eta_{1,2}$ are  
everywhere orthogonal\footnote{This is more restrictive 
than requiring that $\eta_{1,2}$ in (\ref{1}) be nowhere-parallel.}. 
In addition, we look for solutions with nonzero Romans' mass. 
We reformulate the supersymmetry conditions in terms of 
$SU(2)$ structures in section \ref{analysis}. In search for explicit 
solutions we make some further simplifying assumptions; namely, 
we set all nonscalar (in the sense of irreducible 
$SU(2)$ representations) fluxes to zero and we take the dilaton to be 
constant. This is a consistent truncation which, however, turns out to be 
too stringent: as we will see there do not exist 
any supersymmetric vacua of this type. 

A related $\mathcal{N}=1$ type IIA vacuum was constructed in 
\cite{bc}. In that case the manifold $X_6$ was taken to be 
conformally $R^6$ and therefore noncompact, allowing for 
non-constant harmonic functions. 
Taking $X_6$ to be 
$T^6$ instead in the solution of \cite{bc}, would imply that 
the warp factors and the dilaton are constant 
and that the Romans' mass and all the fluxes are zero. 
The solution would therefore degenerate to $R^{1,3}\times T^6$, 
in agreement with our conclusion above. Type IIA supersymmetric 
compactifications to Minkowski space on manifolds of $SU(2)$ 
structure have 
also been considered in \cite{gg2}. 
However, all ten-dimensional vacua in that  
paper  
arise upon reduction of eleven-dimensional supergravity 
solutions and are therefore unrelated to the present 
work\footnote{Recall that for nonzero mass parameter, as is 
the case here, Romans' supergravity 
has no Poincar\'e-invariant lift to eleven dimensions; see 
\cite{hs} for a recent discussion.}.

In section \ref{pri} we proceed to examine 
the case of $\mathcal{N}=2$ supersymmetric (warped) $AdS_4$ vacua. 
Rather than considering the most general spinor Ansatz, 
we will take the 
two Majorana-Weyl supersymmetry parameters $\e_{1,2}$ 
to be of the form
\beal
\e_1&=\theta_+\otimes(\alpha\eta_{1+}+\beta\eta_{1-})+{\rm c.c.}\nn\\
\e_2&=\theta_+\otimes(\gamma\eta_{2+}+\delta\eta_{2-})+{\rm c.c.}~,
\label{1.2}
\end{align}
where $\alpha$, $\beta$, $\gamma$, $\delta$, are functions on $X_6$ 
and $\eta_{1,2}$ are globally-defined, unimodular spinors 
on $X_6$. In addition we  take $\eta_{1,2}$ to be 
orthogonal to each other. Requiring $M_{1,3}$ to be $AdS_4$ implies  
$\alpha=\beta$, $\gamma=\delta$. 
As we show in section \ref{analysis}, 
this requirement is again too stringent and 
in fact there do not exist 
any $\mathcal{N}=2$ IIA vacua of this type. 

Note that 
for an admissible vacuum, except 
for the supersymmetry conditions we also require the 
supergravity equations of 
motion and the Bianchi identities for the forms to be 
satisfied. More generally, 
the no-go theorems of this paper could be by-passed by 
introducing additional sources, for example orientifolds, 
which would modify the equations-of-motion. Alternatively 
one may consider singular 
and/or noncompact `internal' manifolds, higher-order 
stringy corrections, etc. We also emphasize that, 
due to the technical complexity of the task,  
we have not been able to analyze the most general spinor Ansatz 
leading to $SU(2)$ structures. 
We hope to report on that in the future.

The outline of the 
remainder of the paper is as follows: In the next section we review some 
useful facts about (Romans') IIA supergravity. Our Ansatz for the 
ten-dimensional 
$\mathcal{N}=1$ background and the corresponding 
reduction of the supersymmetry conditions are considered in section 
\ref{reduction}. 
Section \ref{koko} contains a brief review 
of $SU(3)$ and $SU(2)$ structures in six dimensions.  
The main analysis of our Ansatz for 
$\mathcal{N}=1$ supersymmetric vacua is contained 
in section \ref{analysis}. Section \ref{pri} 
contains the analysis of our $\mathcal{N}=2$ Ansatz. 
Most of the technical details 
and some further useful formul{\ae} are relegated to the four appendices.

\section{Massive IIA}
\label{kiki}

For the sake of completeness, in this section we note some known facts about
Romans'  
ten-dimensional supergravity. Our notation and conventions are as
in \cite{lt} to which the reader is referred for further details.

The equations of motion for the bosonic fields of massive IIA supergravity are
\cite{roma}
\beal
0&=R_{MN}-\frac{1}{2}\nabla_M\phi\nabla_N\phi-\frac{1}{12}e^{\phi/2}
G_{MPQR}G_{N}{}^{PQR}+\frac{1}{128}e^{\phi/2}g_{MN}G^2\nn\\
&~~~~~-\frac{1}{4}e^{-\phi}
H_{MPQ}H_{N}{}^{PQ}
+\frac{1}{48}e^{-\phi}g_{MN}H^2\nn\\
&~~~~~-2m^2e^{3\phi/2}
B'_{MP}B'_{N}{}^{P}+\frac{m^2}{8}e^{3\phi/2}g_{MN}(B')^2
-\frac{m^2}{4}e^{5\phi/2}g_{MN}\\
0&=\nabla^2\phi-\frac{1}{96}e^{\phi/2}G^2+\frac{1}{12}e^{-\phi}H^2
-\frac{3m^2}{2}e^{3\phi/2}(B')^2-5m^2e^{5\phi/2}\\
\la{heq}
0&=d(e^{-\phi}*H)-\frac{1}{2}G\wedge G+2m~e^{\phi/2}B'\wedge*G
+4m^2e^{3\phi/2} *B'\\
\la{geq}
0&=d(e^{\phi/2}{}*G)-H\wedge G~.
\end{align}
In addition, the forms obey the Bianchi identities
\beal
dB'&=H\nn\\
dH&=0\nn\\
dG&=2mB'\wedge H~.
\label{bianchi1}
\end{align}
To make contact with the massless IIA supergravity of \cite{gp, cw, hn}
one introduces a St\"{u}ckelberg gauge  potential $A$,
with field strength $F=dA$, so that
\beal
m B'&=m B+\frac{1}{2}F~.
\label{masslesslimit}
\end{align}
In the massless limit, $m\longrightarrow 0$, we have $mB'\longrightarrow\frac{1}{2}F$.

\subsection*{Supersymmetry}

The gravitino and dilatino supersymmetry variations read
\beal
\delta\Psi_M={\mathcal{D}_M}\epsilon~
\la{gravitino}
\end{align}
and
\beal
\de\lambda=\Big\{
-\frac{1}{2}\G^M\nabla_M\phi&-\frac{5m~e^{5\phi/4}}{4}
+\frac{3m~e^{3\phi/4}}{8}B'_{MN}\G^{MN}\G_{11}\nn\\
&+\frac{e^{-\phi/2}}{24}H_{MNP}\G^{MNP}\G_{11}
-\frac{e^{\phi/4}}{192}G_{MNPQ}\G^{MNPQ}
\Big\}\e~,
\la{dilatino}
\end{align}
where
\beal
{\mathcal{D}_M}&:=\Big\{\nabla_M-\frac{m~e^{5\phi/4}}{16}\Gamma_M
-\frac{m~e^{3\phi/4}}{32}
B'_{NP}(\G_M{}^{NP}-14\de_M{}^N\G^P)\G_{11}\nn\\
&+\frac{e^{-\phi/2}}{96}H_{NPQ}(\G_M{}^{NPQ}-9\de_M{}^N\G^{PQ})\G_{11}
+\frac{e^{\phi/4}}{256}G_{NPQR}(\G_M{}^{NPQR}
-\frac{20}{3}\de_M{}^N\G^{PQR})\Big\}~.
\end{align}
One can transform to the string frame
by rescaling $e_A{}^M\rightarrow e^{\phi/4}e_A{}^M$.

\subsection*{Integrability}

It was shown in \cite{lt} that imposing supersymmetry
together with the equations of motion for the forms implies
the dilaton equation and the
Einstein equation $E_{MN}=0$, provided $E_{M0}=0$ 
for $M\neq 0$ \footnote{Similar integrability
conditions were derived in \cite{paki, banda, bandb} 
in the context of 
eleven-dimensional supergravity. 
See also \cite{orti} for a recent general discussion.}.

\section{$\mathcal{N}=1$ $M_{1,3}\times_{\omega} X_6$ backgrounds}
\label{reduction}

Let us now assume that spacetime
is of the form of a warped product
$M_{1,3}\times_{\omega} X_6$, where $M_{1,3}$ is
Minkowski or $AdS_4$ and $X_6$ is a compact manifold.
The ten dimensional metric reads
\be
g_{MN}(x,y) =
\left(
\begin{array}{cc}
\Delta^2(y) \hat{g}_{\mu\nu}(x) & 0 \\
0 & \rho_{mn}(y)
\end{array}
\right)
~,
\label{lkpk}
\end{equation}
where $x$ is a coordinate on $M_{1,3}$ and $y$ is a coordinate on $X_6$.
We will also assume that the forms have nonzero
$y$-dependent components
along the internal directions, except for the four-form which will be
allowed to have an additional component proportional to the volume
of $M_{1,3}$
\be
G_{\mu\nu\kappa\lambda}=\sqrt{g_{4}}f(y)\varepsilon_{\mu\nu\kappa\lambda}~,
\label{pioui}
\end{equation}
where $f$ is a real scalar function on $X_6$.
Note that with these assumptions
the $E_{M0}=0$ for $M\neq 0$ condition is satisfied, and therefore
we need only check supersymmetry
the Bianchi identities and the equations of motion for the forms.

\subsection{Massive $\mathcal{N}=1$ vacua and $SU(2)$ structure}
\label{supersymmetry}

On $M_{1,3}$ there is a pair of Weyl spinors (related by complex conjugation),
each of which
satisfies the Killing equation
\be
\hat{\nabla}_\mu\theta_+=W\hat{\g}_\mu\theta_-~;
~~~~~~~
\hat{\nabla}_\mu\theta_-= W^*\hat{\g}_\mu\theta_+ ~,
\la{ads}
\end{equation}
where hatted quantities are computed using
the metric $\hat{g}_{\mu\nu}$, and
the complex constant $W$ is related to the scalar curvature $\hat{R}$
of $M_{1,3}$ through $\hat{R}=-24|W|^2$. 
The reader is referred to \cite{lt} for further details on our spinor conventions
in four, six, ten  dimensions.

It can been shown (see for example \cite{grana} for a recent discussion) that 
the requirement of $\mathcal{N}=1$ supersymmetry\footnote{This corresponds
to four real supercharges; in the present paper we are counting supersymmetries
according to four-dimensional conventions.} 
implies that the  Majorana-Weyl supersymmetry parameter $\e$ is 
decomposed 
under $Spin(1,9)\longrightarrow Spin(1,3)\times Spin(6)$ as
\beal
\e&=\alpha(y)\theta_+\otimes\eta_{1+}+\delta(y) 
\theta_+\otimes\eta_{2-}+{\rm \
c.c.}~,
\label{sp1}
\end{align}
where $\alpha$, $\delta$, are complex functions on $X_6$ and 
$\eta_{1,2}$ is a pair of globally-defined, 
{\it nowhere-vanishing} 
Weyl spinors on $X_6$. Moreover, $\eta_{1,2-}$ are related 
to $\eta_{1,2+}$ by complex 
conjugation. Without loss of generality,
we can choose $\eta_{1,2}$ to be of unit norm. 
In keeping with four-dimensional supersymmetry nomenclature,
we take $\theta$ ($\eta_{1,2}$) to be anticommuting (commuting). 

There are three cases 
according to the relation between $\eta_1$ and $\eta_2$: a) 
$\eta_2$ is everywhere parallel to $\eta_1$ and $X_6$ is of $SU(3)$ 
structure, b) $\eta_2$ is nowhere parallel to $\eta_1$ and $X_6$ is 
of $SU(2)$ structure, or c) at generic points in $X_6$ $\eta_{1,2}$ are 
linearly independent, but there exist points where 
$\eta_2$ becomes parallel to $\eta_1$. The latter case imposes 
no additional topological requirement on $X_6$ other than that 
its structure group should reduce to $SU(3)$. 
 
In the present paper we will take 
$\eta_2$ to be  everywhere {\it orthogonal} to $\eta_1$: 
$\eta_{2+}^+\eta_{1+}=0$. 
This is a special sub-case of b) above and therefore 
$X_{6}$ must be a manifold of $SU(2)$ structure.  
As we will see later in section \ref{analysis}, 
requiring in addition that the Romans' mass be nonzero 
implies that up to a choice of phase
which can be absorbed in the normalization of the spinors $\eta_{1,2}$,
\beal
|\alpha|=\alpha=\delta~.
\end{align}

\subsection{Reduction of the supersymmetry conditions}
\label{roup}

Substituting the spinor Ansatz (\ref{sp1}) in the supersymmetry
transformations we obtain
\beal
0&=\alpha\nabla_m\eta_{1+}
+\partial_m\alpha\eta_{1+}
+\alpha\frac{e^{-\phi/2}}{96}H_{npq}
(\g_m{}^{npq}-9\delta_m{}^n\g^{pq})\eta_{1+}
-\delta\frac{me^{5\phi/4}}{16}\g_m\eta_{2-}\nn\\
&+3i\delta f\frac{e^{\phi/4}}{32}\g_m\eta_{2-}
+\delta\frac{me^{3\phi/4}}{32}B'_{np}
(\g_m{}^{np}-14\delta_m{}^n\g^p)\eta_{2-}\nn\\
&+\delta\frac{e^{\phi/4}}{256}G_{npqr}(\g_m{}^{npqr}-\frac{20}{3}\delta_m{}^n
\g^{pqr})\eta_{2-}
\la{3.6}\\
0&=\delta^*\nabla_m\eta_{2+}
+\partial_m\delta^*\eta_{2+}
-\delta^*\frac{e^{-\phi/2}}{96}H_{npq}
(\g_m{}^{npq}-9\delta_m{}^n\g^{pq})\eta_{2+}
+\alpha^*\frac{me^{5\phi/4}}{16}\g_m\eta_{1-}\nn\\
&+3i\alpha^* f\frac{e^{\phi/4}}{32}\g_m\eta_{1-}
+\alpha^*\frac{me^{3\phi/4}}{32}B'_{np}
(\g_m{}^{np}-14\delta_m{}^n\g^p)\eta_{1-}\nn\\
&-\alpha^*\frac{e^{\phi/4}}{256}G_{npqr}(\g_m{}^{npqr}-\frac{20}{3}\delta_m{}^n
\g^{pqr})\eta_{1-}~,
\la{3.7}
\end{align}
from the `internal' components of the gravitino variation and
\beal
0&=\alpha\Delta^{-1}W\eta_{1+}+\delta^*\frac{me^{5\phi/4}}{16}\eta_{2+}
-5i\delta^*f\frac{e^{\phi/4}}{32}\eta_{2+}
-\delta^*\frac{me^{3\phi/4}}{32}B'_{mn}\g^{mn}\eta_{2+}\nn\\
&+\alpha^*\frac{e^{-\phi/2}}{96}H_{mnp}\g^{mnp}\eta_{1-}
-\delta^*\frac{e^{\phi/4}}{256}G_{mnpq}\g^{mnpq}\eta_{2+}
-\frac{1}{2}\alpha^*\partial_m({\rm ln}\Delta)\g^m\eta_{1-}
\la{3.8}\\
0&=\delta^*\Delta^{-1}W^*\eta_{2+}+\alpha\frac{me^{5\phi/4}}{16}\eta_{1+}
+5i\alpha f\frac{e^{\phi/4}}{32}\eta_{1+}
+\alpha\frac{me^{3\phi/4}}{32}B'_{mn}\g^{mn}\eta_{1+}\nn\\
&+\delta\frac{e^{-\phi/2}}{96}H_{mnp}\g^{mnp}\eta_{2-}
-\alpha\frac{e^{\phi/4}}{256}G_{mnpq}\g^{mnpq}\eta_{1+}
+\frac{1}{2}\delta\partial_m({\rm ln}\Delta)\g^m\eta_{2-}~,
\la{3.9}
\end{align}
from the noncompact piece. Note that these equations are complex.
Similarly from the dilatino we obtain
\beal
0&=\frac{1}{2}\alpha^*\partial_m\phi \g^m  \eta_{1-} -
\alpha^*\frac{e^{-\phi/2}}{24}H_{mnp}\g^{mnp}\eta_{1-}
-\delta^*\frac{5me^{5\phi/4}}{4}\eta_{2+}\nn\\
& +i\delta^* f\frac{e^{\phi/4}}{8}\eta_{2+}
-\delta^*\frac{3me^{3\phi/4}}{8}B'_{mn}\g^{mn}\eta_{2+}
-\delta^*\frac{e^{\phi/4}}{192}G_{mnpq}\g^{mnpq}\eta_{2+}
\la{3.10}\\
0&=\frac{1}{2}\delta\partial_m\phi \g^m  \eta_{2-} +
\delta\frac{e^{-\phi/2}}{24}H_{mnp}\g^{mnp}\eta_{2-}
+\alpha\frac{5me^{5\phi/4}}{4}\eta_{1+}\nn\\
& +i\alpha f\frac{e^{\phi/4}}{8}\eta_{1+}
-\alpha\frac{3me^{3\phi/4}}{8}B'_{mn}\g^{mn}\eta_{1+}
+\alpha\frac{e^{\phi/4}}{192}G_{mnpq}\g^{mnpq}\eta_{1+}~.
\la{3.11}
\end{align}

\section{$SU(2)$ reduction}
\label{koko}

The analysis of the conditions for a supersymmetric vacuum and the characterization
of the solutions 
is greatly facilitated by using the machinery of $G$-structures \cite{gaun}.
The existence of two globally-defined nowhere-vanishing orthogonal spinors $\eta_{1,2}$,
as is the case here, implies the reduction of the structure group of $X_6$ to $SU(2)$. This allows
us to decompose all tensors on $X_6$ in terms of irreducible $SU(2)$ representations.
In the following two subsections we review some of the relevant facts about $SU(3)$ and
$SU(2)$ structures, before we turn to the analysis of the conditions for an $\mathcal{N}=1$
supersymmetric vacuum in section \ref{analysis}.
The details  of the
$SU(2)$ decomposition of the antisymmetric forms of IIA supergravity and the
$SU(2)$ decomposition of the supersymmetry conditions are
given in appendices \ref{decomposition}, \ref{supersymmetryreduction} respectively.

\subsection{$SU(3)$ structure}

The existence of a
nowhere-vanishing globally-defined spinor $\eta_1$ allows us to define the bilinears
\be
J_{mn}:=i\eta^+_{1-}\g_{mn}\eta_{1-} = -i\eta^+_{1+}\g_{mn}\eta_{1+}
\la{j}
\end{equation}
\be
\Omega_{mnp}:=\eta_{1-}^+\g_{mnp}\eta_{1+}; ~~~~~~
\Omega^*_{mnp}=-\eta_{1+}^+\g_{mnp}\eta_{1-}~.
\la{o}
\end{equation}
Note that $J_{mn}$ thus defined is real and
$\Omega$ ($\Omega^*$) is imaginary (anti-) self-dual.
\be
\Omega_{mnp}=\frac{i}{6}\sqrt{\rho_6}~\varepsilon_{mnpijk}\Omega^{ijk}~.
\end{equation}
We choose to  normalize
\be
\eta_{1+}^+\eta_{1+}=\eta_{1-}^+\eta_{1-}=1~.
\la{n}
\end{equation}
Using (\ref{fierz}) one can prove
that $J$, $\Omega$ satisfy
\be
J_m{}^nJ_n{}^p=-\delta_m{}^p
\end{equation}
\be
(\Pi^+)_m{}^n\Omega_{npq}=\Omega_{mpq}; ~~~~~~
(\Pi^-)_m{}^n\Omega_{npq}=0  ~,
\end{equation}
where
\be
(\Pi^\pm)_m{}^n:=\frac{1}{2}(\delta_m{}^n\mp i J_m{}^n)
\la{projectors}
\end{equation}
are the projection operators onto the holomorphic/antiholomorphic parts.
In other words, $J$ defines an almost complex structure
with respect to which $\Omega$ is $(3,0)$.
Moreover (using (\ref{fierz}) again) it follows that
\beal
\Omega\wedge J&=0\nn\\
\Omega\wedge\Omega^*&=\frac{4i}{3}J^3~.
\end{align}
Therefore $J$, $\Omega$, completely specify an $SU(3)$ structure on $X_6$.

Further useful relations can be found in \cite{lt}.

\subsection{$SU(2)$ structure}
\label{structure}

The existence of two orthogonal unimodular globally-defined spinors
$\eta_1$, $\eta_2$ on $X_6$ allows us to define two distinct $SU(3)$ structures
\beal
J_{mn}:=i\eta^+_{1-}\g_{mn}\eta_{1-} ; ~~~~~~
\Omega_{mnp}:=\eta_{1-}^+\g_{mnp}\eta_{1+}
\end{align}
and
\beal
J'_{mn}:=i\eta^+_{2-}\g_{mn}\eta_{2-} ; ~~~~~~
\Omega'_{mnp}:=\eta_{2-}^+\g_{mnp}\eta_{2+}~.
\end{align}
Each of these satisfies all the properties of $SU(3)$ structures
given in the preceding section. It can be shown,
using the above definitions and the Fierz identities in appendix \ref{fierzidentities},  that
\beal
J&=-\frac{i}{2}K\wedge K^*+\widetilde{J}\nn\\
J'&=-\frac{i}{2}K\wedge K^*-\widetilde{J}~,
\end{align}
where
\beal
K_{m}&:=\eta_{2-}^+\g_{m}\eta_{1+}
\end{align}
and
\beal \iota_K\widetilde{J}=0~.
\end{align}
The complex vector $K$ satisfies
\beal
K_mK^m=0; ~~~~~~ K^*_mK^m=2
\end{align}
and is holomorphic with respect to $J$,
\beal
(\Pi^+)_m{}^nK_n=K_m; ~~~~~~ (\Pi^-)_m{}^nK_n=0~.
\end{align}
It follows that the two-form $\widetilde{J}$ is $(1,1)$ with
respect to the almost complex structure $J$.

The two holomorphic three-forms can be expressed
in the following way
\beal
\Omega&=-iK\wedge \omega\nn\\
\Omega'&=iK\wedge \omega^*~,
\end{align}
where
\beal \omega_{mn}&:=i\eta_{1-}^+\g_{mn}\eta_{2-}
\end{align}
satisfies
\beal \iota_K\omega=\iota_{K^*}\omega=0
\end{align}
and is holomorphic with respect to $J$,
\beal (\Pi^+)_m{}^n\omega_{np}=\omega_{mp}; ~~~~~~
(\Pi^-)_m{}^n\omega_{np}=0~.
\end{align}
It is straightforward to show that $\widetilde{J}$, $\omega$
specify an $SU(2)$ structure. Indeed, it follows from the above
formul{\ae} that
\beal
\widetilde{J}&\wedge\omega=0\nn\\
\omega\wedge\omega^*&=2\widetilde{J}\wedge\widetilde{J}~.
\end{align}
The complex vector $K$ specifies an almost product structure
\beal
R_m{}^n:=K_mK^{*n}+ K^*_mK^{n}-\delta_m{}^n~,
\end{align}
such that
\beal
R_m{}^nR_{n}{}^p=\delta_m{}^p~.
\end{align}

Further useful relations are given in appendix \ref{sustructure}.

\section{Analysis of the conditions}
\label{analysis}

To analyze the supersymmetry conditions of section \ref{roup} 
it is useful to note that equations (\ref{3.6}, \ref{3.7})
can be cast in the form
\beal
U_m\eta_{1+}+U_{mn}\g^n\eta_{1-}=0~,
\end{align}
whereas equations (\ref{3.8}-\ref{3.11}) can be written as
\beal
V\eta_{1+}+V_{m}\g^m\eta_{1-}=0~.
\label{pou}
\end{align}
The explicit expressions for the $U$'s and $V$'s
can be read off from the expressions in appendix \ref{supersymmetryreduction}.
The tensors $U$, $V$ further decompose into directions parallel and perpendicular
to the complex vector $K$ defined in \ref{structure}. The components perpendicular to
$K$ are further decomposed in terms of irreducible $SU(2)$ representations.
The details of the decomposition are given in appendix \ref{decomposition}. To illustrate
the procedure, let us decompose
\beal
V_m=v_m+vK_m~,
\end{align}
where $K^mv_m=K^{*m}v_m=0$.
We also noted that
in the decomposition of $V_m$ there are no
terms proportional to $K_m^{*}$, due to \ref{kri}.  It follows that the scalar content
of (\ref{pou}) is equivalent to:
\beal
V=v=0~.
\end{align}
We proceed similarly for all other representations.

Let us consider the scalar component of the supersymmetry equations first.
It is straightforward to show that if there exists a point
$y_0$ in $X_6$ such that $|\alpha(y_0)|\neq |\delta(y_0)|$, 
equations  (\ref{3.8}-\ref{3.11}) 
imply that $m=W=0$. I.e. the mass 
parameter vanishes and the space $M_{1,3}$ reduces to
Minkowski. We would like to look for massive solutions of IIA and hence, up to phases which
can be absorbed in the normalizations of $\eta_{1,2}$ we can take:
\beal
|\alpha|=\alpha=\delta~,
\label{ui}
\end{align}
at each point in $X_6$. 
Let us first analyze the supersymmetry equations (\ref{3.8}-
\ref{3.11}), 
considering each irreducible $SU(2)$ representation in turn. 
The decompositions of all antisymmetric 
tensors in terms of irreducible $SU(2)$ representations 
can be found in appendix \ref{decomposition}. 
One can show that the solution is equivalent to the following conditions:

{\it The \bf{1}}

\beal
f&=0\nn\\
W&=\frac{i\Delta}{8}(me^{3\phi/4}b_2^*+\frac{i}{12}e^{\phi/4}g_2^*)~,
\label{jia}
\end{align}
\beal
mb_1&=0\nn\\
me^{3\phi/4}b_3&=\frac{i}{2}(d_{K}\phi_--d_{K^*}\phi_-)~,
\end{align}
where $d_{K}:=K^m\partial_m$, $d_{K^*}:=K^{*m}\partial_m$ and  
$\phi_{\pm}:=\phi\pm 4{\rm ln}\Delta$. Also
\beal
e^{\phi/4}g_1&=16(me^{5\phi/4}+\frac{1}{4}d_{K}\phi_-+\frac{1}{4}d_{K^*}\phi_-)\nn\\
g_3&=0~,
\end{align}
\beal
e^{-\phi/2}h_1=e^{-\phi/2}h_2^*&=\frac{9me^{3\phi/4}}{2}b_2-\frac{ie^{\phi/4}}{8}g_2\nn\\
e^{-\phi/2}h_3&=12i(me^{5\phi/4}+\frac{1}{8}d_K\phi_-+\frac{1}{4}d_{K^*}\phi_+)~.
\label{5.9}
\end{align}

{\it The \bf{2}}

\beal
\widetilde{\partial}^+_{m}{\rm ln}\Delta&
=\frac{1}{4}\widetilde{\partial}^+_{m}\phi
-\frac{me^{3\phi/4}}{8}\omega_{m}{}^n \wb_{1n}^*
-\frac{e^{\phi/4}}{64}\wg_{2m}^*~,
\end{align}
where $\widetilde{\partial}_m$ is defined in (\ref{yut}) and
$\widetilde{\partial}^{\pm}_m:=(\widetilde{\Pi}^{\pm})_m{}^n
\widetilde{\partial}_n$.
Moreover
\beal
me^{3\phi/4}\widetilde{b}_{2m}&=-me^{3\phi/4}\wb_{1m}^*
-\frac{e^{\phi/4}}{32}\omega^*_{m}{}^n(\wg_{1n}-\wg_{2n}^*)~,
\end{align}
\beal
e^{-\phi/2}\widetilde{h}_{1m}&=\frac{e^{\phi/4}}{4}(\wg_{1m}-\wg_{2m}^*)\nn\\
e^{-\phi/2}\widetilde{h}_{2m}&=-3i\Big\{ 
2\omega_{m}{}^n\widetilde{\partial}^-_{n}\phi
+3me^{3\phi/4}\wb_{1m}+\frac{e^{\phi/4}}{16}\omega_{m}{}^n(\frac{1}{2}\wg^*_{1n}-\wg_{2n})
\Big\}~,
\end{align}

{\it The \bf{3}}

This representation drops out of equations (\ref{3.8}-\ref{3.11}). 

Next we turn to the equations (\ref{3.6},\ref{3.7}).
The fact that $\eta_{1,2}$ are unimodular implies $\nabla(\eta_{1}^+\eta_{1})=0$
and $\nabla(\eta_{2}^+\eta_{2})=0$ which,
taking (\ref{3.6},\ref{3.7}) into account, can be seen
to be equivalent to
\beal
\alpha={\rm constant}\times \Delta^{ -1/2}~.
\end{align}
In addition, the orthogonality of $\eta_{1,2}$ implies 
$\nabla(\eta_{1}^+\eta_{2})=0$ which,
taking (\ref{3.6},\ref{3.7}) into account, leads
to the condition
\beal
h_1=h^*_2=\frac{ie^{3\phi/4}}{4}g_2~.
\end{align}
Comparing with (\ref{5.9}, \ref{jia}) 
we conclude that $mb_2=ie^{-\phi/2}g_2/12$ and $W=0$. 
Note that the $\bf{2}$ representation drops out of the 
orthogonality constraints.

To summarize the conditions so far:
\beal
f, ~W&=0
\label{arcoona}
\end{align}
In
addition, in form notation, 
\beal
mB'&=
\Big[ \frac{im}{4}{\rm Im}( b_1')
+\frac{1}{64}e^{-\phi/2}(\wg_1-\wg_2^*)\Big]\wedge K+{\rm c.c.}
\nn\\
&+m\widetilde{b} -\frac{e^{-3\phi/4}}{4}
\wj ~{\rm Im}(d_{K}\phi_-) +\frac{e^{-\phi/2}}{48}{\rm Im}(\omega g^*_2)\nn\\
H&=\frac{1}{3}\Big\{
\wh+\frac{e^{3\phi/4}}{16}{\rm Im}(\omega g^*_2)
+3ie^{\phi/2}\wj\Big[
me^{5\phi/4}+\frac{1}{8}d_{K}\phi_-+\frac{1}{4}d_{K^*}\phi_+
\Big]
\Big\}\wedge K+{\rm c.c.}\nn\\
&-\frac{e^{3\phi/4}}{16}\wj\wedge {\rm Im}(\wg_1+\wg_2)+2iIm\Big\{
e^{\phi/2}d^+\phi+\frac{3e^{5\phi/4}}{8}b_1'
+\frac{e^{3\phi/4}}{32}(\frac{1}{2}\wg_1
+\wg_2)\Big\}\wedge K\wedge K^*\nn\\
G&=\Big[me^{\phi}+\frac{e^{-\phi/4}}{2}{\rm Re}(
d_K\phi_-)\Big]\wj\wedge\wj 
\nn\\
&-\frac{i}{32}\Big\{(\wg_1-\wg_2)\wedge\wj\wedge K+{\rm c.c.}\Big\}
-\frac{i}{12}\Big[\wg+\frac{1}{4}{\rm Re}(\omega g_2^*)\Big]\wedge K\wedge K^*~,
\end{align}
where we have defined $b_{1m}':=\omega^*_m{}^nb_{1n}$.
The differential equations (\ref{3.6}, \ref{3.7}) determine the 
specific $SU(2)$ structure of $X_6$ and impose further constraints. 
In order to search for explicit solutions we will now make some simplifying assumptions. 
Namely, we will assume that only scalar fluxes are present (i.e. only the 
$\bf{1}$ components of the form fields are nonzero) and that the dilaton is constant
\footnote{In the remainder of this section 
we will absorb all dilaton dependence by a field redefinition 
of the forms.}. 
We use (\ref{3.6}, \ref{3.7}) 
to read off the exterior derivatives on $\Omega$, $J$ and $K$:
\beal
d\Omega&=-i(m\omega+\frac{g_2}{48}\wj)\wedge K\wedge K^*+\frac{g_2}{12}\wj\wedge \wj\nn\\
dJ&=-\frac{1}{24}{\rm Re}(\omega g_2^*)\wedge {\rm Re}(K)\nn\\
dK&=m K\wedge K^*+\frac{i}{24}{\rm Re}(\omega g_2^*)~.
\end{align}
Moreover, we can read off the action of the 
exterior derivative on the $SU(2)$ structure
\beal
d\wj&=0\nn\\
d\omega&=\frac{g_2}{48}\wj\wedge K^*~.
\end{align}
Combining all the above, we note that the nilpotency of the exterior derivative 
$d^2=0$ implies 
\beal
g_2=0~.
\end{align}
It is now straightforward to see that the Bianchi 
identities 
imply that all fluxes are zero and $m=0$, 
contrary to our assumption. We therefore conclude that there are 
no solutions obeying our 
simplified Ansatz.

\section{$\mathcal{N}=2$ $AdS_4$  vacua and $SU(2)$ structure}
\label{pri}

In this section we will search for $\mathcal{N}=2$ supersymmetric 
vacua of the type $AdS_4\times_{\omega} X_6$. In order to simplify the 
computation, we will not consider the most 
general spinor Ansatz. Instead we demand that the 
background be invariant under two supersymmetries $\e_{1,2}$ 
of the form 
\beal
\e_1&=\alpha(y)\theta_+\otimes\eta_{1+}+\beta(y) \theta_+\otimes\eta_{1-}+{\rm c.c.}~,
\label{spinoransatz1}
\end{align}
and 
\beal
\e_2&=\gamma(y)\theta_+\otimes\eta_{2+}+\delta(y) \theta_+\otimes\eta_{2-}+{\rm\
 c.c.}~,
\label{spinoransatz2}
\end{align}
where $\alpha$, $\beta$, $\gamma$, $\delta$, 
are complex functions on $X_6$ and 
$\eta_{1,2}$ are globally-defined, unimodular spinors 
on $X_6$. In addition we will take $\eta_{1,2}$ to be 
orthogonal to each other.  Consequently,  
$X_6$ must be a manifold of $SU(2)$ structure. 
As we will see later in section \ref{an},
under these assumptions  
supersymmetry implies that up to a choice of phase
which can be absorbed in the normalizations of the spinors $\eta_{1,2}$,
\beal
\alpha&=\beta\nn\\
\gamma&=\delta~.
\end{align}

\subsection{Reduction of the supersymmetry conditions}

Substituting the spinor Ansatz (\ref{spinoransatz1}) in the supersymmetry
transformations we obtain
\beal
0&=\alpha\nabla_m\eta_{1+}
+\partial_m\alpha\eta_{1+}
+\alpha\frac{e^{-\phi/2}}{96}H_{npq}
(\g_m{}^{npq}-9\delta_m{}^n\g^{pq})\eta_{1+}
-\beta\frac{me^{5\phi/4}}{16}\g_m\eta_{1-}\nn\\
&+3i\beta f\frac{e^{\phi/4}}{32}\g_m\eta_{1-}
+\beta\frac{me^{3\phi/4}}{32}B'_{np}
(\g_m{}^{np}-14\delta_m{}^n\g^p)\eta_{1-}\nn\\
&+\beta\frac{e^{\phi/4}}{256}G_{npqr}(\g_m{}^{npqr}-\frac{20}{3}\delta_m{}^n
\g^{pqr})\eta_{1-}
\la{eva}\\
0&=\beta^*\nabla_m\eta_{1+}
+\partial_m\beta^*\eta_{1+}
-\beta^*\frac{e^{-\phi/2}}{96}H_{npq}
(\g_m{}^{npq}-9\delta_m{}^n\g^{pq})\eta_{1+}
+\alpha^*\frac{me^{5\phi/4}}{16}\g_m\eta_{1-}\nn\\
&+3i\alpha^* f\frac{e^{\phi/4}}{32}\g_m\eta_{1-}
+\alpha^*\frac{me^{3\phi/4}}{32}B'_{np}
(\g_m{}^{np}-14\delta_m{}^n\g^p)\eta_{1-}\nn\\
&-\alpha^*\frac{e^{\phi/4}}{256}G_{npqr}(\g_m{}^{npqr}-\frac{20}{3}\delta_m{}^n
\g^{pqr})\eta_{1-}~,
\la{duo}
\end{align}
from the `internal' components of the gravitino variation and
\beal
0&=\alpha\Delta^{-1}W\eta_{1+}+\beta^*\frac{me^{5\phi/4}}{16}\eta_{1+}
-5i\beta^*f\frac{e^{\phi/4}}{32}\eta_{1+}
-\beta^*\frac{me^{3\phi/4}}{32}B'_{mn}\g^{mn}\eta_{1+}\nn\\
&+\alpha^*\frac{e^{-\phi/2}}{96}H_{mnp}\g^{mnp}\eta_{1-}
-\beta^*\frac{e^{\phi/4}}{256}G_{mnpq}\g^{mnpq}\eta_{1+}
-\frac{1}{2}\alpha^*\partial_m({\rm ln}\Delta)\g^m\eta_{1-}
\la{tria}\\
0&=\beta^*\Delta^{-1}W^*\eta_{1+}+\alpha\frac{me^{5\phi/4}}{16}\eta_{1+}
+5i\alpha f\frac{e^{\phi/4}}{32}\eta_{1+}
+\alpha\frac{me^{3\phi/4}}{32}B'_{mn}\g^{mn}\eta_{1+}\nn\\
&+\beta\frac{e^{-\phi/2}}{96}H_{mnp}\g^{mnp}\eta_{1-}
-\alpha\frac{e^{\phi/4}}{256}G_{mnpq}\g^{mnpq}\eta_{1+}
+\frac{1}{2}\beta\partial_m({\rm ln}\Delta)\g^m\eta_{1-}~,
\la{tessera}
\end{align}
from the noncompact piece. Note that these equations are complex.
Similarly from the dilatino we obtain
\beal
0&=\frac{1}{2}\alpha^*\partial_m\phi \g^m  \eta_{1-} -
\alpha^*\frac{e^{-\phi/2}}{24}H_{mnp}\g^{mnp}\eta_{1-}
-\beta^*\frac{5me^{5\phi/4}}{4}\eta_{1+}\nn\\
& +i\beta^* f\frac{e^{\phi/4}}{8}\eta_{1+}
-\beta^*\frac{3me^{3\phi/4}}{8}B'_{mn}\g^{mn}\eta_{1+}
-\beta^*\frac{e^{\phi/4}}{192}G_{mnpq}\g^{mnpq}\eta_{1+}
\la{pevte}\\
0&=\frac{1}{2}\beta\partial_m\phi \g^m  \eta_{1-} +
\beta\frac{e^{-\phi/2}}{24}H_{mnp}\g^{mnp}\eta_{1-}
+\alpha\frac{5me^{5\phi/4}}{4}\eta_{1+}\nn\\
& +i\alpha f\frac{e^{\phi/4}}{8}\eta_{1+}
-\alpha\frac{3me^{3\phi/4}}{8}B'_{mn}\g^{mn}\eta_{1+}
+\alpha\frac{e^{\phi/4}}{192}G_{mnpq}\g^{mnpq}\eta_{1+}~.
\la{e3i}
\end{align}
A second set of conditions follows from the second supersymmetry
(\ref{spinoransatz2}). These can be obtained from the ones above by
substituting $(\alpha, \beta,\eta_1)\longrightarrow (\gamma,\delta,\eta_2)$.

\subsection{Analysis of the conditions}
\label{an}

Let us consider the scalar component of the supersymmetry equations first.
It is straightforward to show that if there exists a point
$y_0$ in $X_6$ such that $|\alpha(y_0)|\neq |\beta(y_0)|$ or  $|\gamma(y_0)|\neq |\delta(y_0)|$,
equations  (\ref{tria}-\ref{e3i})
(and the ones obtained from them
by substituting $(\alpha, \beta,\eta_1)\longrightarrow (\gamma,\delta,\eta_2)$)
imply that $m,~f,~W=0$. I.e. the space $M_{1,3}$ reduces to
Minkowski, which is contrary to our assumption. Hence, up to phases which
can be absorbed in the normalizations of $\eta_{1,2}$ we can take:
\beal
\alpha&=\beta\nn\\
\gamma&=\delta~,
\label{uib}
\end{align}
at each point in $X_6$. Taking (\ref{uib}) into account, it is useful to
note that the
supersymmetry conditions (\ref{eva}, \ref{duo}),
as well as the ones obtained from them
by substituting $(\alpha, \eta_1)\longrightarrow (\gamma,\eta_2)$,
are equivalent to the following
set of equations:
\beal
0&=\nabla_m\eta_{1+}
+\partial_m{\rm ln}|\alpha|\eta_{1+}+3if\frac{e^{\phi/4}}{32}\g_m\eta_{1-}
+\frac{me^{3\phi/4}}{32}B'_{np}
(\g_m{}^{np}-14\delta_m{}^n\g^p)\eta_{1-}
\la{evaprime}\\
0&=\partial_m{\rm ln}\Big(\frac{\alpha}{|\alpha|}\Big)\eta_{1+}
+\frac{e^{-\phi/2}}{96}H_{npq}
(\g_m{}^{npq}-9\delta_m{}^n\g^{pq})\eta_{1+}
-\frac{me^{5\phi/4}}{16}\g_m\eta_{1-}\nn\\
&+\frac{e^{\phi/4}}{256}G_{npqr}(\g_m{}^{npqr}-\frac{20}{3}\delta_m{}^n
\g^{pqr})\eta_{1-}~
\la{duoprime}
\end{align}
and 
\beal
0&=\nabla_m\eta_{2+}
+\partial_m{\rm ln}|\gamma|\eta_{2+}+3if\frac{e^{\phi/4}}{32}\g_m\eta_{2-}
+\frac{me^{3\phi/4}}{32}B'_{np}
(\g_m{}^{np}-14\delta_m{}^n\g^p)\eta_{2-}
\la{evaprimeprime}\\
0&=\partial_m{\rm ln}\Big(\frac{\gamma}{|\gamma|}\Big)\eta_{2+}
+\frac{e^{-\phi/2}}{96}H_{npq}
(\g_m{}^{npq}-9\delta_m{}^n\g^{pq})\eta_{2+}
-\frac{me^{5\phi/4}}{16}\g_m\eta_{2-}\nn\\
&+\frac{e^{\phi/4}}{256}G_{npqr}(\g_m{}^{npqr}-\frac{20}{3}\delta_m{}^n
\g^{pqr})\eta_{2-}~.
\la{duoprimeprime}
\end{align}
Let us first analyze the supersymmetry equations (\ref{duoprime},
\ref{duoprimeprime}), (\ref{tria}-\ref{e3i}) and the ones obtained from them
by $(\alpha, \eta_1)\longrightarrow (\gamma,\eta_2)$,
considering each irreducible $SU(2)$ representation in turn. 
The decompositions of all antisymmetric 
tensors in terms of irreducible $SU(2)$ representations 
can be found in appendix \ref{decomposition}. 
One can show that the solution is equivalent to the following conditions:

{\it The $\bf{1}$}

\beal
m&=0\nn\\
W&=\frac{i\Delta}{6}\Big(\frac{\alpha}{|\alpha|}\Big)^{-2}fe^{\phi/4}~.
\label{ji}
\end{align}
\beal
mb_1&=\frac{f}{6}e^{-\phi/2}\nn\\
mb_2&=\frac{4i}{3}e^{-3\phi/4}d_K\phi\nn\\
mb_3&=0~,
\end{align}
\beal
g_1=g_2=g_3=0~,
\end{align}
\beal
h_1=h_2=h_3=0~,
\end{align}
\beal
\frac{\alpha}{|\alpha|}&=\pm\frac{\gamma}{|\gamma|}\nn\\
d_K\Big(\frac{\alpha}{|\alpha|}\Big)
=&d_{K^*}\Big(\frac{\alpha}{|\alpha|}\Big)=0~,
\end{align}
\beal
d_{K}{\rm ln}\Delta&=-\frac{1}{12}d_{K}\phi\nn\\
d_{K}\phi&=d_{K^*}\phi~.
\label{ty}
\end{align}
The first line of (\ref{ty}) together with the second line of (\ref{ji})
and the fact that $W$ is constant, imply:
\beal
d_{K}{\rm ln}f=-\frac{1}{6}d_{K}\phi~.
\end{align}

{\it The $\bf{2}$}

\beal
m\widetilde{b}_{1m}&=\frac{4i}{3}e^{-3\phi/4}\widetilde{\partial}^{+}_m\phi\nn\\
m\widetilde{b}_{2m}&=-\frac{4i}{3}e^{-3\phi/4}\widetilde{\partial}^{-}_m\phi~,
\end{align}
\beal
\widetilde{h}_{1m}=\widetilde{h}_{2m}=0~,
\end{align}
\beal
\widetilde{g}_{1m}=\widetilde{g}_{2m}=0~,
\end{align}
\beal
\widetilde{\partial}_m\Big(\frac{\alpha}{|\alpha|}\Big)&=0\nn\\
\widetilde{\partial}_m{\rm ln}\Delta&=-\frac{1}{12}\widetilde{\partial}_m\phi~.
\end{align}

{\it The $\bf{3}$}

\beal
\widetilde{h}_{mn}=\widetilde{g}_{mn}=0~.
\end{align}
The relations derived so far imply $\Delta={\rm constant}\times e^{-\phi/12}$,
$f={\rm constant}\times e^{-\phi/6}$, as well as $H=0$, $G=fdVol_4$,
where $dVol_4$ is the volume element of $M_{1,3}$ in the warped metric.
It then follows from the Bianchi identity (\ref{bianchi1}) for the $G$ field that
$\phi={\rm constant}$.

To summarize the conditions so far:
\beal
m&=0\nn\\
W&=\frac{i\Delta}{6}\Big(\frac{\alpha}{|\alpha|}\Big)^{-2}fe^{\phi/4}\nn\\
\frac{\gamma}{|\gamma|}&=\pm \frac{\alpha}{|\alpha|}\nn\\
\frac{\alpha}{|\alpha|}, ~\Delta, ~\phi, ~f&={\rm constant}~.
\label{arcoonab}
\end{align}
In
addition, in form notation, 
\beal
F&=
\widetilde{f}-\frac{i}{6}f e^{-\phi/2}K\wedge K^*
\nn\\
H&=0\nn\\
G&=fdVol_4~.
\end{align}
Note that we have taken (\ref{masslesslimit})
and the fact that $m=0$ into account, and we have set
$m\widetilde{b}_{mn}=\frac{1}{2}\widetilde{f}_{mn}$.

Next we turn to the equations (\ref{evaprime},\ref{evaprimeprime}).
The fact that $\eta_{1,2}$ are unimodular implies $\nabla(\eta_{1}^+\eta_{1})=0$
and $\nabla(\eta_{2}^+\eta_{2})=0$ which,
taking (\ref{evaprime},\ref{evaprimeprime}) into account, can be seen
to be equivalent to $|\alpha|,~|\gamma|={\rm constant}$. Together
with (\ref{arcoonab}) this implies
\beal
\alpha,~\gamma={\rm constant}~.
\end{align}
In addition, the orthogonality of $\eta_{1,2}$ implies $\nabla(\eta_{1}^+\eta_{2})=0$ which,
taking (\ref{evaprime},\ref{evaprimeprime}) into account, leads
to the condition
\beal
f=0~.
\end{align}
Taking (\ref{arcoonab}) into account, this implies $W=0$ and $M_{1,3}$ reduces
to Minkowski space\footnote{It is not difficult to see that 
in addition the equations of motion impose $\widetilde{f}=0$ 
and therefore all fluxes are zero.}. 
Of course, this is contrary to our assumption of a
(warped) $AdS_4$ vacuum. We are
therefore led to the conclusion that
there are no $\mathcal{N}=2$ solutions of type IIA supergravity satisfying 
our requirements.

\vskip20pt

\noindent{\bf Acknowledgements:} We are grateful to C. Jeschek
for valuable discussions. This work is supported in part by the
EU-RTN network {\sl Constituents, Fundamental Forces and Symmetries
of the Universe} (MRTN-CT-2004-005104).

\appendix

\section{Fierz identities}
\label{fierzidentities}

Using definitions
(\ref{j},\ref{o})
we find
\beal
\eta_{1-}^\alpha\eta_{1+}^\beta &=\frac{1}{4}(P_- C^{-1})^{\alpha\beta}
+\frac{i}{8}J_{mn}(P_- \g^{mn}C^{-1})^{\alpha\beta}\nn\\
\eta_{1+}^\alpha\eta_{1+}^\beta &=-\frac{1}{48}
\Om_{mnp}(P_+ \g^{mnp}C^{-1})^{\alpha\beta}\nn\\
\eta_{1-}^\alpha\eta_{1-}^\beta &=\frac{1}{48}
\Om^*_{mnp}(P_- \g^{mnp}C^{-1})^{\alpha\beta}~
\la{fierz}
\end{align}
and similarly for $\eta_2\otimes\eta_2$, by replacing
$(J,\Omega)\rightarrow (J',\Omega')$. Moreover for $\eta_1\otimes\eta_2$ we have
\beal
\eta_{1-}^\alpha\eta_{2+}^\beta &=\frac{i}{8}\omega^*_{mn}(P_- \g^{mn}C^{-1})^{\alpha\beta}\nn\\
\eta_{2-}^\alpha\eta_{1+}^\beta &=\frac{i}{8}\omega_{mn}(P_- \g^{mn}C^{-1})^{\alpha\beta}\nn\\
\eta_{1+}^\alpha\eta_{2+}^\beta &=\frac{1}{4}
K_m(P_+ \g^{m}C^{-1})^{\alpha\beta}-\frac{1}{48}
\widetilde{\Om}_{mnp}(P_+ \g^{mnp}C^{-1})^{\alpha\beta}\nn\\
\eta_{1-}^\alpha\eta_{2-}^\beta &=-\frac{1}{4}
K_m^*(P_- \g^{m}C^{-1})^{\alpha\beta}
+\frac{1}{48}
\widetilde{\Om}^*_{mnp}(P_- \g^{mnp}C^{-1})^{\alpha\beta}~,
\la{fierzdou}
\end{align}
where
\beal
\widetilde{\Om}_{mnp}&:=(\eta_{2-}^+\g_{mnp}\eta_{1+})~.
\end{align}
The latter is imaginary self-dual
\beal
\widetilde{\Om}_{mnp}=\frac{i}{6}\sqrt{\rho_6}~\varepsilon_{mnpqrs}\widetilde{\Om}^{qrs}
\end{align}
and obeys
\beal
\widetilde{\Om}\wedge\omega=\widetilde{\Om}\wedge\omega^*=0~.
\end{align}
As follows from (\ref{fierzdou}), the two globally defined spinors are related via
\beal
\eta_{2+}=-\frac{1}{2}K^m\g_{m}\eta_{1-}~.
\label{secspi}
\end{align}
We also note the following relations,
\beal
0&=(\Pi^+)_m{}^n\gamma_n\eta_{1-}\nn\\
\gamma_{mn}\eta_{1+}&=iJ_{mn}\eta_{1+}+\frac{1}{2}\Omega_{mnp}\gamma^p\eta_{1-}\nn\\
\gamma_{mnp}\eta_{1-}&=-3iJ_{[mn} \gamma_{p]}\eta_{1-}-\Omega^*_{mnp}\eta_{1+}~.
\label{kri}
\end{align}
A useful formula following from (\ref{secspi}, \ref{kri}) is
\beal
\g^m\eta_{2-}=K^{*m}\eta_{1+}+\frac{i}{2}\omega_{mn}\g^n\eta_{1-}~.
\end{align}

\section{$SU(2)$ structure}
\label{sustructure}

Here we give some further useful relations pertaining to the $SU(2)$ structure. 

It follows from (\ref{kri}) that
\beal
0&=(\widetilde{\Pi}^+)_m{}^n\gamma_n\eta_{1-}~,
\end{align}
where
\beal
(\widetilde{\Pi}^{\pm})_{mk}:=\frac{1}{2}(\widetilde{\rho}_{mk}\mp
i\widetilde{J}_{mk})~
\end{align}
and
\beal
\widetilde{\rho}_{mk}:=\rho_{mk}-\frac{1}{2}(K_mK^{*}_{k}+ K^*_mK_{k})~.
\end{align}
Note that
\beal
K^m\widetilde{\rho}_{mk}=0~.
\end{align}
Some further useful identities are
\beal 
\widetilde{J}_{mn}\widetilde{J}^{n}{}_{k}&=-\widetilde{\rho}_{mk}\nn\\
\widetilde{J}_{m}{}^n\omega_{nk}&=i\omega_{mk}\nn\\
\omega_{mn}\omega^{*nk}
&=-4(\widetilde{\Pi}^+)_{m}{}^k\nn\\
\omega_{mn}\omega^{*ij}
&=8(\widetilde{\Pi}^+)_{[m}{}^i(\widetilde{\Pi}^+)_{n]}{}^j\nn\\
(\widetilde{\Pi}^+)_{m}{}^k&=(\Pi^+)_{m}{}^k-\frac{1}{2}K_mK^{*k}~.
\end{align}

\section{$SU(2)$ tensor decompositions}
\label{decomposition}

In terms of the $SU(2)$ structure, the form fields of IIA supergravity decompose as follows.

{\it Two-form}
\beal
B'_{mn}=b_{mn}+b_{[m}K_{n]}+b^*_{[m}K^*_{n]}+ib_1K_{[m}K^*_{n]}~,
\end{align}
where
\beal
K^ib_{im}=K^ib_{i}=K^{*i}b_{i}=0
\end{align}
and
\beal
K^{i}B'_{im}&=-b^*_m-ib_1K_m \nn\\
K^iK^{*j}B'_{ij}&=-2ib_1~.
\end{align}
Note that $b_1$ is real.
We can further decompose
\beal
b_{mn}=\widetilde{b}_{mn}+\frac{1}{8}\omega^*_{mn}b_2+\frac{1}{8}\omega_{mn}b^*_2
+\frac{1}{4}\widetilde{J}_{mn}b_3~,
\end{align}
where $\widetilde{b}_{mn}$ is $(1,1)$ and traceless with respect
to $\widetilde{J}_{mn}$, i.e. it transforms in the $\bf{3}$ of
$SU(2)$. The scalar $b_2$ is complex whereas $b_3$ is real. We
have
\beal
b_2&=\omega^{mn}b_{mn}\nn\\
b_3&=\widetilde{J}^{mn}b_{mn}~.
\end{align}
Finally,
\beal b_m=-\frac{1}{4}\omega^*_{m}{}^i\widetilde{b}_{1i}
-\frac{1}{4}\omega_{m}{}^i\widetilde{b}_{2i},
\end{align}
where $(\Pi^-)_m{}^n\widetilde{b}_{1n}= (\Pi^+)_m{}^n\widetilde{b}_{2n}=0$. Both 
$\widetilde{b}_{1i}$, $\widetilde{b}_{2i}$ transform in the ${\bf{2}}$ 
of $SU(2)$. We have
\beal
\widetilde{b}_{1i}&=\omega_m{}^nb_n\nn\\
\widetilde{b}_{2i}&=\omega^*_m{}^{n}b_n~.
\end{align}

{\it Three-form}
\beal
H_{mnp}=h_{mnp}+h_{[mn}K_{p]}+h^*_{[mn}K^*_{p]}+ih_{[m}K_{n}K^*_{p]}~,
\end{align}
where
\beal
K^ih_{imn}=K^ih_{im}=K^{*i}h_{im}=K^ih_{i}=0
\end{align}
and
\beal
K^{i}H_{imn}&=\frac{2}{3}h^*_{mn}+\frac{2i}{3}h_{[m}K_{n]}\nn\\
K^iK^{*j}H_{ijm}&=-\frac{2i}{3}h_m~.
\end{align}
Note that $h_m$ is real whereas $h_{mn}$ is complex.
We can further decompose
\beal
h_{mnp}=-\frac{3}{32}\omega_{[mn}\omega^*_{p]}{}^{i}\widetilde{h}_{1i}
-\frac{3}{32}\omega^*_{[mn}\omega_{p]}{}^i\widetilde{h}^*_{1i},
\end{align}
where $(\Pi^-)_m{}^n\widetilde{h}_{1n}=0$. We have
\beal \widetilde{h}_{1m}&=\omega_m{}^i\omega^{*jk}h_{ijk}~.
\end{align}
Moreover
\beal
h_{mn}=\widetilde{h}_{mn}+\frac{1}{8}\omega^*_{mn}h_1+\frac{1}{8}\omega_{mn}h_2
+\frac{1}{4}\widetilde{J}_{mn}h_3~,
\end{align}
where $\widetilde{h}_{mn}$ is complex and it is $(1,1)$ and
traceless with respect to $\widetilde{J}_{mn}$. The scalars
$h_{1,2,3}$ are complex. We have
\beal
h_1&=\omega^{mn}h_{mn}\nn\\
h_2&=\omega^{*mn}h_{mn}\nn\\
h_3&=\widetilde{J}^{mn}h_{mn}~.
\end{align}
Finally,
\beal h_m=-\frac{1}{4}\omega^*_{m}{}^i\widetilde{h}_{2i}
-\frac{1}{4}\omega_{m}{}^i\widetilde{h}^*_{2i},
\end{align}
where $(\Pi^-)_m{}^n\widetilde{h}_{2n}=0$. We have
\beal \widetilde{h}_{2i}&=\omega_m{}^nh_n~.
\end{align}

{\it Four-form}
\beal
G_{mnpq}=g_{mnpq}+g_{[mnp}K_{q]}+g^*_{[mnp}K^*_{q]}+ig_{[mn}K_{p}K^*_{q]}~,
\end{align}
where
\beal
K^ig_{imnp}=K^ig_{imn}=K^{*i}g_{imn}=K^ig_{im}=0
\end{align}
and
\beal
K^{i}G_{imnp}&=-\frac{1}{2}g^*_{mnp}-\frac{i}{2}g_{[mn}K_{p]}\nn\\
K^iK^{*j}G_{ijmn}&=-\frac{i}{3}g_{mn}~.
\end{align}
Note that $g_{mnpq}$, $g_{mn}$ are real whereas $g_{mnp}$ is complex.
We can further decompose
\beal
g_{mnpq}=\frac{3}{8}\widetilde{J}_{[mn}\widetilde{J}_{pq]}g_1,
\end{align}
where the scalar $g_1$ is real. We have
\beal g_1=\widetilde{J}^{mn}\widetilde{J}^{pq}g_{mnpq}~.
\end{align}
Moreover
\beal
g_{mnp}=-\frac{3}{32}\omega_{[mn}\omega^*_{p]}{}^i\widetilde{g}_{1i}
-\frac{3}{32}\omega^*_{[mn}\omega_{p]}{}^i\widetilde{g}_{2i},
\end{align}
where $(\Pi^-)_m{}^n\widetilde{g}_{1n}=(\Pi^+)_m{}^n\widetilde{g}_{2n}=0$. We have
\beal
\widetilde{g}_{1m}&=\omega_m{}^i\omega^{*jk}g_{ijk}\nn\\
\widetilde{g}_{2m}&=\omega^*_m{}^i\omega^{jk}g_{ijk} ~.
\end{align}
Finally,
\beal
g_{mn}=\widetilde{g}_{mn}+\frac{1}{8}\omega^*_{mn}g_2+\frac{1}{8}\omega_{mn}g^*_2
+\frac{1}{4}\widetilde{J}_{mn}g_3~,
\end{align}
where $\widetilde{g}_{mn}$ is real and it is traceless with
respect to $\widetilde{J}_{mn}$. The scalar $g_{2}$ is complex
whereas $g_{3}$ is real. We have
\beal
g_2&=\omega^{mn}g_{mn}\nn\\
g_3&=\widetilde{J}^{mn}g_{mn}~.
\end{align}

\section{$SU(2)$ supersymmetry reduction}
\label{supersymmetryreduction}

Using (\ref{secspi}) and the decompositions of section  \ref{decomposition},
it follows that conditions (\ref{eva}, \ref{duo}) can be cast in the form
\beal
U_m\eta_{1+}+U_{mn}\g^n\eta_{1-}=0~,
\end{align}
whereas conditions (\ref{tria}-\ref{e3i}) can be written as
\beal
V\eta_{1+}+V_{m}\g^m\eta_{1-}=0~,
\end{align}
for some $U_m$, $U_{mn}$, $V$, $V_{m}$. The explicit expressions for the $U$'s and $V$'s
can be readily read off from the following decompositions in terms of 
irreducible $SU(2)$ representations:

{\it Two-form}

\beal
(\g_m{}^{np}B'_{np}&-14\g^{p}B'_{mp})\eta_{1-}=\Big\{
-iK^*_mb^*_2-2i\wb_{2m}
\Big\}\eta_{1+}\nn\\
&+\Big\{ (2b_1-3b_3)\widetilde{J}_{mn} -14\wb_{mn}
-2i\widetilde{J}_m{}^i\wb_{in}-\frac{3}{2}b_2^*\omega_{mn}\nn\\
&-\frac{3}{2}K_m\omega_n{}^i\wb_{2i}-2K^*_m\omega_n{}^i\wb^*_{1i}
+K_n({2}\omega^*_m{}^i\wb_{1i}+\frac{3}{2}\omega_m{}^i\wb_{2i})
+iK_nK_m^*(7b_1-\frac{1}{2}b_3) \Big\}\g^n\eta_{1-} ~,
\end{align}
\beal (\g_m{}^{np}B'_{np}&-14\g^{p}B'_{mp})\eta_{2-}=\Big\{
iK^*_m(14b_1+b_3)+3\omega^*_m{}^i\wb_{1i}+4\omega_m{}^i\wb_{2i}
\Big\}\eta_{1+}\nn\\
&+\Big\{-3b_2\widetilde{J}_{mn}
+(b_1+\frac{3}{2}b_3)\omega_{mn}-i\omega_m{}^i\wb_{in}
-7i\omega_n{}^i\wb_{im}\nn\\
&-3iK_m\wb_{1n}-4iK^*_m\wb^*_{2n}
-{i}K_n\wb_{1m}
-\frac{i}{2}b_2 K_nK_m^*
\Big\}\g^n\eta_{1-}
\end{align}
and
\beal
B'_{mn}\g^{mn}\eta_{1+}=i(b_3+2b_1)\eta_{1+}
+\Big\{-i\wb^*_{2m}-\frac{i}{2}b_2K_m\Big\}\g^m\eta_{1-}
~,
\end{align}
\beal B'_{mn}\g^{mn}\eta_{2+}=ib_2^*\eta_{1+}
+\Big\{\frac{1}{2}\omega_m{}^i\wb_{1i}^*+iK_m(\frac{1}{2}b_3-b_1)\Big\}\g^m\eta_{1-}~.
\end{align}

{\it Three-form}

\beal H_{npq}(\g_m{}^{npq}-&9\delta_m{}^n\g^{pq})\eta_{1+}= \Big\{
-3\widetilde{h}_{1m}+\frac{3}{2}\widetilde{h}_{1m}^*
+2i\omega_m{}^i\widetilde{h}_{2i}^*
+i\omega^*_m{}^i\widetilde{h}_{2i}-4ih_3K_m-2ih_3^*K_m^* \Big\}\eta_{1+}\nn\\
&+\Big\{3K_m\widetilde{h}_{2n}+K_n\wh_{2m}-\frac{3i}{8}K_m\omega_n{}^i\wh_{1i}^*
-\frac{9i}{8}K_n\omega_m{}^i\wh_{1i}^*+2ih_1K_mK_n+ih_2^*K_nK_m^*\nn\\
&-2h_2^*~\widetilde{J}_{mn}+h_3^*\omega_{mn}
-2i\omega_m{}^j\widetilde{h}^*_{jn}
-6i\omega_n{}{}^j\widetilde{h}^*_{jm} \Big\}\g^n\eta_{1-}~,
\end{align}
\beal H_{npq}(\g_m{}^{npq}-&9\delta_m{}^n\g^{pq})\eta_{2+}= \Big\{
-2\widetilde{h}_{2m}^*
+\frac{9i}{4}\omega^*_m{}^i\widetilde{h}_{1i}
-4iK_mh_2-2iK_m^*h_1^*\Big\}\eta_{1+}\nn\\
&+\Big\{\frac{3}{2}K_n\wh_{1m}^*-\frac{3}{4}K_n\wh_{1m}
-\frac{3}{4}K_m\widetilde{h}_{1n}+\frac{3i}{2}K_m\omega_n{}^i\wh_{2i}^*
-\frac{i}{2}K_n\omega_m{}^i\wh_{2i}^*-iK_n\omega^*_m{}^i\wh_{2i}
\nn\\
&-2ih_3K_mK_n-ih_3^*K_nK_m^*
+2h_3^*~\widetilde{J}_{mn}+4i\widetilde{J}_m{}^j\widetilde{h}^*_{jn}
+\omega_{mn}{h}^*_{1}+12\wh^*_{mn} \Big\}\g^n\eta_{1-}
\end{align}
and
\beal
H_{mnp}\g^{mnp}\eta_{1-}=-2ih_2\eta_{1+}+\Big\{\frac{i}{2}\omega_m{}^i\wh_{2i}^*
-\frac{3}{4}\wh_{1m}-ih_3K_m \Big\}\g^m\eta_{1-}~,
\end{align}
\beal H_{mnp}\g^{mnp}\eta_{2-}=2ih_3\eta_{1+}
+\Big\{\frac{3i}{8}\omega_m{}^i\wh_{1i}^*
-\wh_{2m}-ih_1K_m\Big\}\g^m\eta_{1-}~.
\end{align}

{\it Four-form}

\beal
(\g_m{}^{npqr}G_{npqr}&-\frac{20}{3}\g^{pqr}G_{mpqr})\eta_{1-}=\Big\{
\frac{10}{3}K^*_mg^*_2-\frac{5i}{2}\omega^*_m{}^i\wg_{1i}
\Big\}\eta_{1+}\nn\\
&+\Big\{i(5g_1+\frac{2}{3}g_3)\widetilde{J}_{mn}
+\frac{20i}{3}\wg_{mn}
-4\widetilde{J}_m{}^i\wg_{in}+\frac{i}{3}g_2^*\omega_{mn}\nn\\
&-\frac{1}{2}K_m\wg_{1n}-2K^*_m\wg^*_{2n}+K_n(\frac{1}{2}\wg_{1m}-2\wg_{2m})
+K_nK_m^*(\frac{5}{3}g_3-\frac{3}{2}g_1) \Big\}\g^n\eta_{1-} ~,
\end{align}
\beal
(\g_m{}^{npqr}G_{npqr}&-\frac{20}{3}\g^{pqr}G_{mpqr})\eta_{2-}=\Big\{
-K^*_m(3g_1+\frac{10}{3}g_3)-4\wg_{1m}+\wg_{2m}
\Big\}\eta_{1+}\nn\\
&+\Big\{\frac{2i}{3}g_2\widetilde{J}_{mn}
+(\frac{5i}{2}g_1-\frac{i}{3}g_3)\omega_{mn}-2\omega_m{}^i\wg_{in}
-\frac{10}{3}\omega_n{}^i\wg_{im}\nn\\
&+\frac{i}{4}K_m\omega_n{}^i\wg_{2i}+iK^*_m\omega_n{}^i\wg^*_{1i}
-\frac{5i}{4}K_n\omega_m{}^i\wg_{2i} +\frac{5}{3}g_2 K_nK_m^*
\Big\}\g^n\eta_{1-}
\end{align}
and
\beal G_{mnpq}\g^{mnpq}\eta_{1+}=-(3g_1+2g_3)\eta_{1+}
+\Big\{\frac{3i}{4}\omega_m{}^i\wg_{1i}^*+g_2K_m\Big\}\g^m\eta_{1-}
~,
\end{align}
\beal
G_{mnpq}\g^{mnpq}\eta_{2+}=-2g_2^*\eta_{1+}
+\Big\{\frac{3}{2}\wg_{2m}^*+K_m(\frac{3}{2}g_1-g_3)\Big\}\g^m\eta_{1-}~.
\end{align}

Finally, a derivative ($\partial_m$) on $X_6$ will be decomposed as
\beal
\partial_m=\widetilde{\partial}_m+\frac{1}{2}K_mK^{*n}\partial_n+
\frac{1}{2}K^*_mK^{n}\partial_n~,
\label{yut}
\end{align}
so that
\beal
\iota_{K}\widetilde{\partial}=0~.
\end{align}
%

%
%


\begin{thebibliography}{99}


\bibitem{gaun}
J.~P.~Gauntlett, D.~Martelli, S.~Pakis and D.~Waldram,
``G-structures and wrapped NS5-branes,''
Commun.\ Math.\ Phys.\  {\bf 247} (2004) 421, hep-th/0205050.



\bibitem{gauntb}
J.~P.~Gauntlett,
``Classifying supergravity solutions,''
Fortsch.\ Phys.\  {\bf 53} (2005) 468, hep-th/0501229.


\bibitem{lt}
D.~L\"{u}st and D.~Tsimpis,
``Supersymmetric AdS(4) compactifications of IIA supergravity,''
JHEP {\bf 0502} (2005) 027, hep-th/0412250.



\bibitem{bca}
K.~Behrndt and M.~Cvetic,
``General N = 1 supersymmetric flux vacua of (massive) type IIA string
theory,'' hep-th/0403049.

\bibitem{bc}
K.~Behrndt and M.~Cvetic,
``General N = 1 supersymmetric fluxes in massive type IIA string theory,'' 
hep-th/0407263.






\bibitem{g1}
J.~Louis and A.~Micu,
``Type II theories compactified on Calabi-Yau threefolds in the presence  of
background fluxes,'' Nucl.\ Phys.\ B {\bf 635} (2002) 395, hep-th/0202168.

\bibitem{g2}
J.~P.~Derendinger, C.~Kounnas, P.~M.~Petropoulos and F.~Zwirner,
``Superpotentials in IIA compactifications with general fluxes,''
Nucl.\ Phys.\ B {\bf 715} (2005) 211, hep-th/0411276.

\bibitem{g3}
S.~Kachru and A.~K.~Kashani-Poor,
``Moduli potentials in type IIA compactifications with RR and NS flux,''
JHEP {\bf 0503} (2005) 066, hep-th/0411279.


\bibitem{g31}
G.~Villadoro and F.~Zwirner,
``N = 1 effective potential from dual type-IIA D6/O6 orientifolds with
general fluxes,'' hep-th/0503169.


\bibitem{g4}
O.~DeWolfe, A.~Giryavets, S.~Kachru and W.~Taylor,
``Type IIA moduli stabilization,'' hep-th/0505160.

\bibitem{g5}
T.~House and E.~Palti,
``Effective action of (massive) IIA on manifolds with SU(3) structure,'' 
hep-th/0505177.

\bibitem{granb}
M.~Grana, J.~Louis and D.~Waldram,
``Hitchin functionals in N = 2 supergravity,'' hep-th/0505264.


\bibitem{Cardoso:2002hd}
G.~L.~Cardoso, G.~Curio, G.~Dall'Agata, 
D.~L\"ust, P.~Manousselis and G.~Zoupanos,
``Non-K\"ahler string backgrounds and their five torsion classes,''
Nucl.\ Phys.\ B {\bf 652}, 5 (2003)
hep-th/0211118.





\bibitem{gg1}
J.~P.~Gauntlett, D.~Martelli and D.~Waldram,
``Superstrings with intrinsic torsion,''
Phys.\ Rev.\ D {\bf 69} (2004) 086002, hep-th/0302158.




\bibitem{gg2}
G.~Dall'Agata and N.~Prezas,
``N = 1 geometries for M-theory and type IIA strings with fluxes,''
Phys.\ Rev.\ D {\bf 69} (2004) 066004, hep-th/0311146.




\bibitem{gg3}
M.~Grana, R.~Minasian, M.~Petrini and A.~Tomasiello,
``Supersymmetric backgrounds from generalized Calabi-Yau manifolds,''
JHEP {\bf 0408} (2004) 046, hep-th/0406137.

\bibitem{gg4}
M.~Grana, R.~Minasian, M.~Petrini and A.~Tomasiello,
``Type II strings and generalized Calabi-Yau manifolds,''
hep-th/0409176.






\bibitem{grana}
M.~Grana, R.~Minasian, M.~Petrini and A.~Tomasiello,
``Generalized structures of N = 1 vacua,'' hep-th/0505212.


























\bibitem{js}
J.~H.~Schwarz,
``Superconformal Chern-Simons theories,''
JHEP {\bf 0411} (2004) 078, hep-th/0411077.

\bibitem{nunez}
C.~Nunez, I.~Y.~Park, M.~Schvellinger and T.~A.~Tran,
``Supergravity duals of gauge theories from F(4) gauged supergravity in  six dimensions,''
JHEP {\bf 0104} (2001) 025, hep-th/0103080.




\bibitem{jesc}
C.~Jeschek,
``Generalized Calabi-Yau structures and mirror symmetry,'' hep-th/0406046. 

\bibitem{kapu}
A.~Kapustin and Y.~Li, 
``Topological sigma-models with H-flux and twisted generalized complex
manifolds,'' hep-th/0407249.


\bibitem{jescb} 
C.~Jeschek and F.~Witt,
``Generalised G(2)-structures and type IIB superstrings,''
JHEP {\bf 0503} (2005) 053, hep-th/0412280.



\bibitem{hs}
P.~S.~Howe and E.~Sezgin,
``The supermembrane revisited,''
Class.\ Quant.\ Grav.\  {\bf 22} (2005) 2167, hep-th/0412245.





\bibitem{roma}
L.~J.~Romans,
``Massive N=2a Supergravity In Ten-Dimensions,''
Phys.\ Lett.\ B {\bf 169} (1986) 374.

\bibitem{gp}
F.~Giani and M.~Pernici,
``N=2 Supergravity In Ten-Dimensions,''
Phys.\ Rev.\ D {\bf 30} (1984) 325.

\bibitem{cw}
I.~C.~G.~Campbell and P.~C.~West,
``N=2 D = 10 Nonchiral Supergravity And Its Spontaneous Compactification,''
Nucl.\ Phys.\ B {\bf 243} (1984) 112.

\bibitem{hn}
M.~Huq and M.~A.~Namazie,
``Kaluza-Klein Supergravity In Ten-Dimensions,''
Class.\ Quant.\ Grav.\  {\bf 2} (1985) 293
[Erratum-ibid.\  {\bf 2} (1985) 597].




\bibitem{paki}
J.~P.~Gauntlett and S.~Pakis,
``The geometry of D = 11 Killing spinors. ((T),''
JHEP {\bf 0304} (2003) 039, hep-th/0212008 .

\bibitem{banda}
I.~A.~Bandos, J.~A.~de Azcarraga, J.~M.~Izquierdo, M.~Picon and O.~Varela,
``On BPS preons, generalized holonomies and D = 11 supergravities,''
Phys.\ Rev.\ D {\bf 69} (2004) 105010, hep-th/0312266.


\bibitem{bandb}
I.~A.~Bandos, J.~A.~de Azcarraga, M.~Picon and O.~Varela,
``Generalized curvature and the equations of D = 11 supergravity,''
Phys.\ Lett.\ B {\bf 615} (2005) 127, hep-th/0501007.


\bibitem{orti}
J.~Bellorin and T.~Ortin,
``A note on simple applications of the Killing spinor identities,''
Phys.\ Lett.\ B {\bf 616} (2005) 118, hep-th/0501246.











%




\end{thebibliography}
\end{document}